\documentclass[twocolumn,prl,showpacs,byrevtex]{revtex4-1}
\usepackage{epsf,graphicx,amssymb,amsfonts,amsmath,amsthm}

\begin{document}
\title{Gravity wave turbulence revealed by horizontal vibrations of the container}
\author{B. Issenmann}
\author{E. Falcon}
\affiliation{Univ Paris Diderot, Sorbonne Paris Cit\'e, MSC, UMR 7057 CNRS, F-75013 Paris, France}

\date{\today}

\begin{abstract}We experimentally study the role of the forcing on gravity-capillary wave turbulence. Previous laboratory experiments using spatially localized forcing (vibrating blades) have shown that the frequency power-law exponent of the gravity wave spectrum depends on the forcing parameters. By horizontally vibrating the whole container, we observe a spectrum exponent that does not depend on the forcing parameters for both gravity and capillary regimes. This spatially extended forcing leads to a gravity spectrum exponent in better agreement with the theory than by using a spatially localized forcing. The role of the vessel shape has been also studied. Finally, the wave spectrum is found to scale linearly with the injected power for both regimes whatever the forcing type used.
\end{abstract}
%Hydrodynamic waves, Nonlinear dynamics and chaos, Chaos in fluid dynamics, Turbulent flows, , 92.10.Lq Turbulence oceanic 

\maketitle

When waves of large enough amplitudes propagate within a dispersive medium, the nonlinear interactions generate waves at different scales. This energy transfer from the large scales (where the energy is injected) to the small scales (where it is dissipated) is called wave turbulence. It occurs in various domains: optical waves, surface or internal waves in oceanography, astrophysical plasma waves, Rossby waves in geophysics, elastic or spin waves in solids (for recent reviews see ~\cite{Falcon2010, Newell2011,Residori2012}). Since the end of the 1960s, weak turbulence theory describes the wave turbulence regimes in almost all fields involving waves~\cite{Zakharov1992}. It assumes strong hypotheses such as those addressing weakly nonlinear, isotropic and homogeneous random waves in an infinite size system with scale separation between injection and dissipation of energy. It notably predicts analytical solutions for the spectrum of a weakly nonlinear wave field at equilibrium or in a stationary out-of equilibrium regime.

While homogeneity and isotropy are two premises of the theory, laboratory experiments generally use spatially localized forcing to generate wave turbulence (e.g., elastic waves on plates, or surface waves on a fluid). The use of a spatially homogeneous forcing is thus of primary interest to probe the validity domain of this theory. Previous experiments were performed by vertically vibrating a vessel filled with a fluid using the Faraday instability to homogeneously generate capillary wave turbulence~\cite{Levinsen00,Snouck2009,Xia2010}. However, this forcing generates localized structures and discrete resonance peaks in the wave spectrum.

In oceans, the gravity wave spectrum depends on numerous forcing parameters (wind, fetch, sea severity, etc). Consequently, in situ spectra are usually fitted with many parameters~\cite{Ochi1998}, and some data are quantitatively in rough agreement with the weak turbulence predictions~\cite{Toba73}. However, recent well controlled laboratory experiments show deviations from the predictions for the scaling of the gravity wave spectrum when waves are locally generated (using vibrating blades at the surface of a fluid)~\cite{Falcon2007, Denissenko2007}. In this case, the exponent of the frequency power-law spectrum of gravity waves depends on the forcing parameters (amplitude and frequency bandwidth)~\cite{Falcon2007, Denissenko2007} instead of being independent, as expected theoretically. The origin of this discrepancy remains an open problem. It has been suggested to be due to finite size effect~\cite{Falcon2007} or to the presence of strong nonlinear waves~\cite{Cobelli2011}.

In this Letter, we study gravity-capillary wave turbulence subjected to horizontal random vibrations of the whole container. The frequency power-laws of the wave spectrum are found independent of the forcing parameters in both gravity and capillary regimes, with a rough agreement with weak turbulence theory.  The probability distribution of the wave height and the scaling of the wave spectrum with the forcing amplitude is also measured. Note that horizontal vibrations of a container at a single frequency have been used to study liquid sloshing motions~\cite{Royon07}, the shape of steep capillary-gravity waves arisen through Kelvin-Helmholtz instability of two immiscible liquids~\cite{Jalikop09}, and in other dissipative systems driven far from equilibrium such as granular materials~\cite{granular}. To our knowledge, no experiment using horizontal random vibrations of the container has been performed so far to study hydrodynamic wave turbulence.

\begin{figure}[t!]
	\includegraphics[width=8cm]{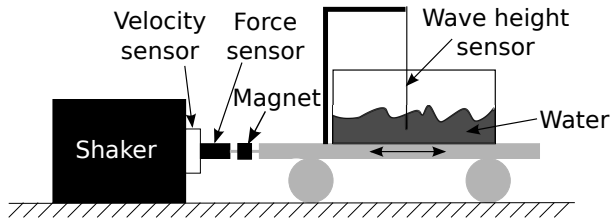} 
	\caption{Experimental setup.}
	\label{IssenmannSetup}
\end{figure}

The experimental setup is shown in Fig.~\ref{IssenmannSetup}. A circular vessel, 22 cm in diameter, is mounted on 4 ball bearing wheels and is horizontally vibrated using an electromagnetic shaker. The container is filled with water up to a depth $h=3$ cm leading to an almost deep water limit ($\lambda \lesssim 2\pi h$ for our range of wavelengths $\lambda$). The shaker (LDS V406/PA 100E) is driven by a random noise forcing low-pass filtered within a frequency bandwidth between $1$~Hz and $f_{p}$ ($f_{p}$ being from $5$ to $7$~Hz). A force sensor (FGP Instr. NTC) is fixed to the shaker axis to measure the instantaneous force $F(t)$ applied by the shaker to the container. The instantaneous velocity $V(t)$ of the container is measured using a home made coil placed on the shaker axis \cite{Falcon2008}. A magnet links the container to the shaker axis (and the axis of both sensors) to impose a force on the container in the direction of the shaker axis. The surface wave height $\eta(t)$ is measured by a home made capacitive sensor \cite{Falcon2007}. This sensor moves together with the vessel, $\eta(t)$ thus being measured in the container framework.  Typical wave mean steepness $s$ ranges from 0.01 to 0.10, estimated as $s=k^*/\sigma_{\eta}$ with $\sigma_{\eta}$ the rms value of $\eta(t)$, and $k*$ the wavenumber of the first normal mode of the vessel (roughly corresponding to the maximum frequency of the spectrum). $F(t)$ and $V(t)$ are recorded for 5~min to compute the mean power $P$ injected to the system (see below). $\eta(t)$ is recorded for 5~min and 30 min, respectively, to compute its power spectrum and its probability distribution. The location of the capacitive sensor has no influence on the spectrum. We are far from conditions of resonance sloshing generating waves strongly coupled with the bulk flow such as swirling waves \cite{Royon07}. Note also that the maximum forcing amplitude is less than the onset of the water drop ejection or wave breaking. 

\begin{figure}[t!]
	\includegraphics[width=8cm]{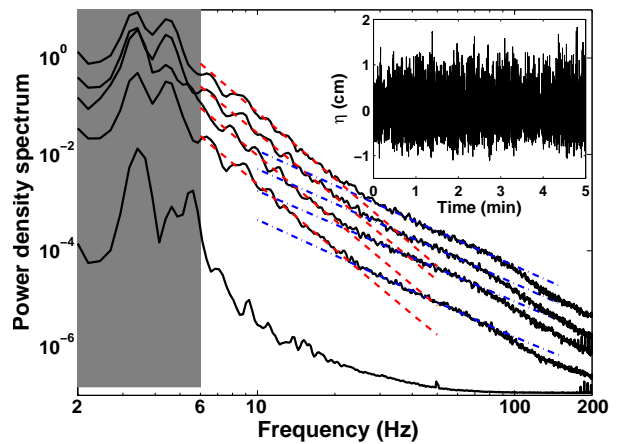} 
	\caption{(Color online) Spectra of the wave height for different injected powers: $P = 1.2$, $14.6$, $17.2$, $23.6$ and $28.5$~mW (bottom to top). Frequency bandwidth of the forcing: 1-6~Hz (colored area). Curves are vertically shifted for clarity. Dashed (red) lines: Power-law fits of the gravity spectra. Dash-dotted (blue) lines: Power-law fits of the capillary spectra. Inset: Typical temporal evolution of the wave height.}
	\label{IssenmannSpectre}
\end{figure}

\begin{figure}[t!]
	\includegraphics[width=8cm]{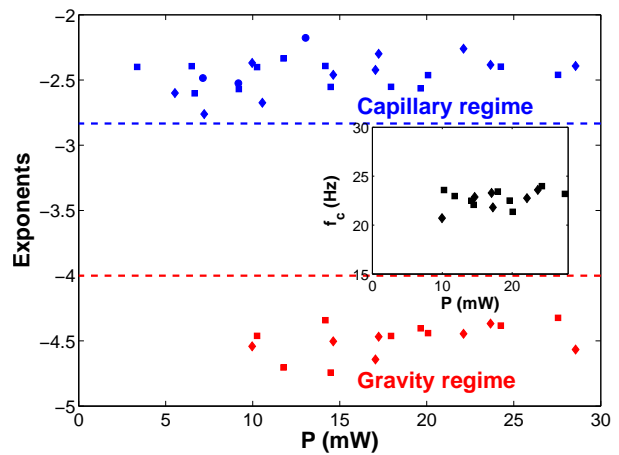} 
	\caption{(Color online) Exponents of the frequency power-law spectra of the capillary and gravity regimes as a function of injected power. Different frequency bandwidths of the forcing: 1 -- 5~Hz ($\blacksquare$), 1 -- 6~Hz ($\blacklozenge$), 1 -- 7~Hz ($\bullet$). Theoretical exponents: $-17/6$ [top (blue) dashed line], and $-4$ [bottom (red) dashed line] for the capillary and gravity wave turbulence regimes. Inset: Crossover frequency between both regimes. Same symbols as in the main figure.}
	\label{IssenmannExposants}
\end{figure}

A typical temporal recording of $\eta(t)$ is shown in the inset of Fig.~\ref{IssenmannSpectre}. $\eta(t)$ displays erratic motion with $\langle \eta \rangle=0$. Its power density spectrum is shown in Fig.~\ref{IssenmannSpectre}. Note that two peaks are visible (at $3.4$ and $4.5\pm 0.2$ Hz) that correspond to theoretical vessel eigenvalue modes \cite{Lamb}. Here, we are interested in the part of the spectrum not directly excited by the forcing ($f > 6$ Hz). At low forcing amplitude, no power law is observed and no wave turbulence regime occurs. At high enough forcing, two frequency-power laws are observed in the spectrum corresponding to the gravity and capillary wave turbulence regimes at low and high frequency, respectively. Similar results were obtained with a vibrating blade forcing \cite{Falcon2007}. The transition between both regimes occurs at a crossover frequency close to $20$~Hz corresponding to $\lambda\approx 1$~cm. The spectrum strongly decreases at high frequency ($\gtrsim 100$ Hz) due to dissipation. When the forcing amplitude is increased, the frequency-power law fits are roughly parallel for each regime. The exponents of the frequency-power laws are shown in Fig.~\ref{IssenmannExposants}. Both gravity and capillary exponents are found independent of the forcing parameters for our range of injected power, taking values of $-4.5\pm 0.2$ and $-2.4\pm 0.3$, respectively. These differ from results of previous studies with localized vibrating blades~\cite{Falcon2007,Denissenko2007} where the gravity spectrum exponent was strongly dependent on the forcing parameters, taking values between $-7$ to $-4$ for the same range of injected power~\cite{Falcon2007}. The exponents obtained here are however slightly different from the theory. Indeed, the gravity exponent is between $-4$ and $-5$. These values correspond  respectively to the weak turbulence spectrum $S_\eta^{grav}(f)\propto \epsilon^{1/3}gf^{-4}$ \cite{Zakharov1967}, and the Phillips spectrum $\propto \epsilon^{0}gf^{-5}$ for sharp crested waves \cite{Phillips1985}, $\epsilon$ being the energy flux, $f$ the frequency, and $g$ the acceleration of gravity. A possible explanation for this deviation is that most of the waves are strongly nonlinear with the wave crest propagating with a preserved shape  \cite{Kuznetsov04}.
 
The spectrum of wave crest ridges having a fractal dimension in the range $0 \leq D<2$ is predicted to scale as $\propto \epsilon^{(2-D)/3}g^{1+D}f^{-3-D}$ \cite{Denissenko2007}. The  gravity spectrum found experimentally in $f^{-4.5}$ thus corresponds to $D=1.5$. The capillary exponent is also found slightly shifted (see Fig.~\ref{IssenmannExposants}) with respect to the weak turbulence prediction $S_\eta^{cap}(f)\propto\epsilon^{1/2}\left(\gamma/\rho\right)^{1/6}f^{-17/6}$ \cite{Zakharov1967-2}, $\gamma$ and $\rho$ being the surface tension and the density of the fluid.

\begin{figure}[t!]
	\includegraphics[width=8.5cm]{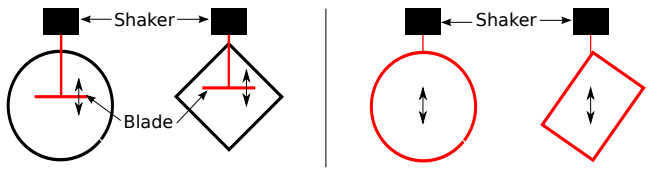} 
	\caption{(Color online) Top view of the tested setups. The shaker horizontally vibrates a blade on the surface of water (left) or the whole container (right) as in Fig.~\ref{IssenmannSetup}. Both  experiments are performed with a circular and a rectangular vessel. The light gray (red) color denotes motion parts, and black color denotes fixed parts.}
	\label{IssenmannSchemas}
\end{figure}

The crossover frequency $f_c$ between both regimes is measured on the spectrum in Fig.~\ref{IssenmannSpectre} as the intersection of both power laws. $f_c$ is shown in the inset of Fig.~\ref{IssenmannExposants} for different forcing parameters. For such a horizontal forcing of the whole container, $f_c$ is roughly independent of the forcing parameters. This result differs from previous studies with a vibrating blade forcing~\cite{Falcon2007} where $f_c$ depended on the forcing parameters in a range from $15$ to $35$~Hz for the same range of injected power and the same container size. 

Consequently, horizontally vibrating the whole container is better than using vibrating blades or parametric forcing to reach a continuum wave turbulence regime independent on the forcing parameters. The main reason is that this forcing is expected to be more spatially homogeneous, and thus better approaches the corresponding theoretical hypothesis even if other assumptions are still not met, like weak nonlinearity and infinite vessel size.

To test the role of the vessel shape and of the type of forcing on gravity-capillary wave turbulence, experiments on two vessels were performed: the circular vessel (22~cm in diameter) and a rectangular one ($15\times 19$~cm$^2$). Two types of forcing were tested for each vessel: a localized vibrating blade and a horizontal forcing of the container as shown on Fig. \ref{IssenmannSchemas}. To avoid the predominance of the eigenvalue frequencies and sloshing modes of the rectangular container, its diagonal is chosen in the same direction as the shaker axis (see Fig. \ref{IssenmannSchemas}). We find that the frequency power-law exponent of the gravity spectrum depends on the forcing parameters: i) with the vibrating blade forcing regardless of the vessel shape, ii) with the rectangular vessel whatever the forcing type. The gravity spectrum exponent is found independent of the forcing parameters only when horizontally vibrating the circular container. 
Although the direction of the forcing is favored in any case, a circular vessel is more isotropic than a rectangular one due to the various wave reflection directions generated by the curved boundary. Thus, beyond homogeneity of the forcing, isotropy is also necessary to reach a gravity spectrum exponent independent of the forcing parameters.

\begin{figure}[t!]
	\includegraphics[width=8cm]{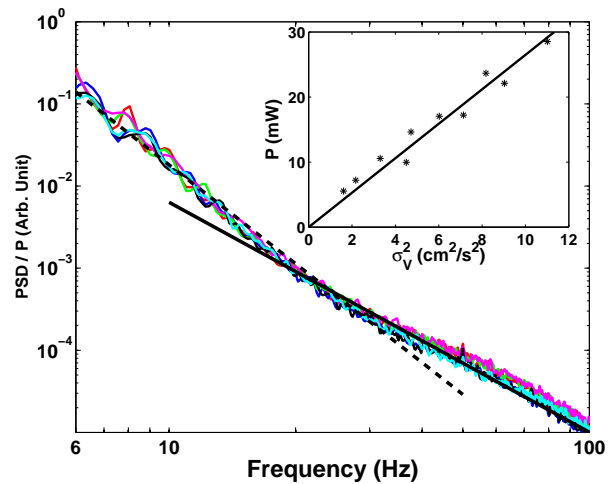} 
	\caption{(Color online) Power spectra of the wave height rescaled by the mean injected power $P$ for $P=10$, 14.6, 17.2, 22.1, 23.6, 28.5~mW. Forcing: 1 -- 6~Hz. Dashed (solid) line is the power-law fit of slope -4 (-2.8) respectively. Inset: $P$ vs. $\sigma_V^2$. Slope of solid line is 2.6 mWs$^2/$cm$^2$. Forcing: 1 -- 6~Hz. }
	\label{SpectreRescalePuissance}
\end{figure}

Let us now focus on the scaling of the wave height spectrum with the mean injected power. The power injected by the shaker to the system corrected by its inertia is $\mathcal{P}(t)=(F - mdV/dt)V$. $m=3.1$~kg is the moving system mass (including the fluid). The mean injected power, $P\equiv \langle \mathcal{P} \rangle$, linearly increases with the variance of the shaker velocity $\sigma_V^2\equiv \langle V^2\rangle$ (see inset of Fig.~\ref{SpectreRescalePuissance}). $\langle \cdot \rangle$ denotes the temporal average. 

The height spectrum is found to scale as $P^{1\pm0.1}$ for both regimes over almost one order of magnitude in $P$ (see Fig.~\ref{SpectreRescalePuissance}). This scaling does not depend on the vessel geometry used. A similar spectrum scaling $\sim P^{1}$ has been observed for both regimes with a vibrating blade forcing \cite{Falcon2007} for the same range of $P$, for the capillary regime with a parametric forcing \cite{Xia2010}, and for the inverse cascade of gravity wave turbulence \cite{Deike11}. This linear scaling is in disagreement with the weak turbulence theory that predicts a spectrum $\sim \epsilon ^{1/3}$ in the gravity regime and $\sim \epsilon ^{1/2}$ in the capillary regime (see above). Experimentally, the mean energy flux $\epsilon$ is usually estimated by the measurement of $P/(\rho \mathcal{S})$,  $\mathcal{S}$ being the immersed moving surface. It is likely that a part of the power is directly provided to the bulk flow and dissipated by viscosity without cascading through the wave field. Although this mechanism is certainly present, it is unlikely to be the dominant one. Indeed, the scaling law of the spectrum with $P$ is the same when the forcing is parametric, or by using wave makers, or by horizontally vibrating the container, while those forcings generate very different bulk flows. We rather think that the part of the injected power directly generating large scale waves only transfers a small amount of energy flux to higher harmonics compared to the direct dissipation of large scale waves by viscosity. This speculation is strengthened by recent experiments of decaying wave turbulence on the surface of a fluid that have shown that only a small part of the initial power injected into the waves feeds the capillary cascade, whereas the major part is dissipated at large scales \cite{Deike12}. This unknown dissipated fraction of injected power could explain the discrepancy with weak turbulence theory for the scaling of the spectrum with $P$. Other possible origins of this discrepancy might be due to finite size effects (by inhibiting the energy transfers among large scale waves) \cite{Falcon2007}, or the presence of strong fluctuations of the injected power \cite{Falcon2008}. 

\begin{figure}[t!]
	\includegraphics[width=8cm]{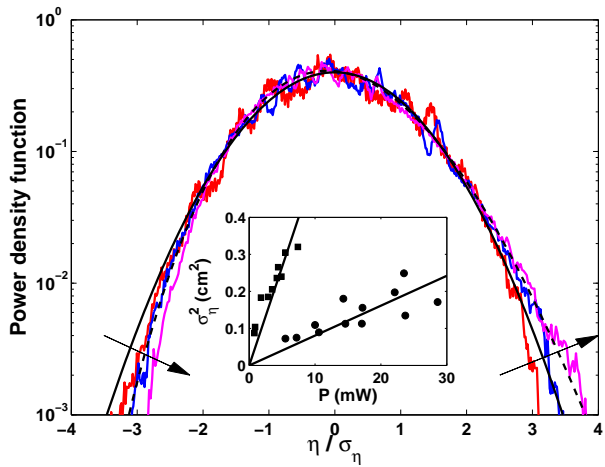} 
	\caption{(Color online) Probability density function of the wave height for $P=1.2$ (red), $2.7$ (blue), $14.8$~mW (magenta) (arrows indicate increasing power) corresponding to $s=$0.018, 0.037 and 0.075. Solid black line: Gaussian with zero mean and unit standard deviation. Dashed black line: Tayfun distribution with $s=0.075$. Forcing: 1 - 6~Hz. Inset:  $\sigma_{\eta}^2$ vs. $P$ for a forcing of the whole container ($\bullet$) or with a vibrating blade ($\blacksquare$). Slopes are respectively $8.1$ and $53$~cm$^2/$W. Forcing: 1 - 6~Hz.}
	\label{IssenmannPDF}
\end{figure}

Finally, the probability density function (PDF) of the wave height normalized by its rms value, $\eta / \sigma_{\eta}$, is shown in Fig.~\ref{IssenmannPDF}. At low forcing, it is symmetric and is roughly fitted by a Gaussian function with zero mean and unit standard deviation. At high enough amplitude, it becomes asymmetric suggesting that large crests are more probable than deep troughs as usual in laboratory experiments with a vibrating blade forcing~\cite{Falcon2007,Onorato2004,Falcon2011} or in oceanography~\cite{Ochi1998,Forristall2000,Socquet2005}. At high forcing, the PDF tends towards a Tayfun distribution (the first non linear correction to the Gaussian) that reads $p[\tilde{\eta}]=\int_0^{\infty}\exp \left( \left[ -x^2-(1-c)^2\right] /(2s^2)\right) /(\pi sc)dx$ where $c=\sqrt{1+2s\tilde{\eta}+x^2}$, $\tilde{\eta}=\eta/\sigma_\eta$, and $s$ the mean wave steepness~\cite{Tayfun1980,Falcon2011}. No adjustable parameter is used here. The shape of PDF($\eta/\sigma_{\eta}$) thus is similar to the one obtained with a vibrating blade forcing. We also find that $\sigma_{\eta}^2=aP$ for both forcing types with different proportionality constants $a$ (see inset of Fig.~\ref{IssenmannPDF}). For a vibrating blade forcing, $P$ was shown to be proportional to $\mathcal{S}$~\cite{Falcon2007}. Here, we have checked that $a\sim 1/\mathcal{S}$ for both methods of forcing. Indeed, the ratio of slopes in the inset of Fig.~\ref{IssenmannPDF} is equal (with a 4\% accuracy) to the inverse of the ratio of the immersed surfaces of the blade and of the container boundary.  Finally, the experimental results, $P \sim \sigma_{\eta}^2$ and $S_{\eta}(f) \sim P^1$, are coherent since by definition $\int_0^{\infty} S_{\eta}(f)df=\sigma_{\eta}^2/(2\pi)$. 

In conclusion, we have introduced a new type of forcing to study gravity-capillary wave turbulence. With this spatially extended forcing, the frequency power-laws of the height spectrum are found independent of the forcing parameters for both gravity and capillary regimes. This contrasts with results of previous experiments using a spatially localized forcing where the gravity spectrum exponent depended on the forcing parameters \cite{Falcon2007,Cobelli2011}.   Our study suggests that the dependence should be related to the inhomogeneity and the anisotropy of the localized forcing. The gravity spectrum exponent found here is slightly different from the one predicted by weak turbulence theory due to the presence of strong nonlinear waves. Finally, an explanation for the discrepancy observed with the theory for the spectrum scaling with $P$ is also given, and applies regardless of the forcing used.

\begin{acknowledgments}
We thank M. Berhanu for fruitful discussions, A. Lantheaume, C. Laroche and J. Servais for technical assistance. B. I. thanks CNRS for funding his postdoctoral research fellow. This work has been supported by ANR Turbulon 12-BS04-0005. 
\end{acknowledgments}

%%%%%%%%%%%%%%%%%%%%%%%%%%%%%%%%%%%%%%
%%%%%%%%%%%% REFERENCES %%%%%%%%%%%%%%%%%%
%%%%%%%%%%%%%%%%%%%%%%%%%%%%%%%%%%%%%%


\begin{thebibliography}{0}%
\makeatletter
\providecommand \@ifxundefined [1]{%
 \@ifx{#1\undefined}
}%
\providecommand \@ifnum [1]{%
 \ifnum #1\expandafter \@firstoftwo
 \else \expandafter \@secondoftwo
 \fi
}%
\providecommand \@ifx [1]{%
 \ifx #1\expandafter \@firstoftwo
 \else \expandafter \@secondoftwo
 \fi
}%
\providecommand \natexlab [1]{#1}%
\providecommand \enquote  [1]{``#1''}%
\providecommand \bibnamefont  [1]{#1}%
\providecommand \bibfnamefont [1]{#1}%
\providecommand \citenamefont [1]{#1}%
\providecommand \href@noop [0]{\@secondoftwo}%
\providecommand \href [0]{\begingroup \@sanitize@url \@href}%
\providecommand \@href[1]{\@@startlink{#1}\@@href}%
\providecommand \@@href[1]{\endgroup#1\@@endlink}%
\providecommand \@sanitize@url [0]{\catcode `\\12\catcode `\$12\catcode
  `\&12\catcode `\#12\catcode `\^12\catcode `\_12\catcode `\%12\relax}%
\providecommand \@@startlink[1]{}%
\providecommand \@@endlink[0]{}%
\providecommand \url  [0]{\begingroup\@sanitize@url \@url }%
\providecommand \@url [1]{\endgroup\@href {#1}{\urlprefix }}%
\providecommand \urlprefix  [0]{URL }%
\providecommand \Eprint [0]{\href }%
\providecommand \doibase [0]{http://dx.doi.org/}%
\providecommand \selectlanguage [0]{\@gobble}%
\providecommand \bibinfo  [0]{\@secondoftwo}%
\providecommand \bibfield  [0]{\@secondoftwo}%
\providecommand \translation [1]{[#1]}%
\providecommand \BibitemOpen [0]{}%
\providecommand \bibitemStop [0]{}%
\providecommand \bibitemNoStop [0]{.\EOS\space}%
\providecommand \EOS [0]{\spacefactor3000\relax}%
\providecommand \BibitemShut  [1]{\csname bibitem#1\endcsname}%
\let\auto@bib@innerbib\@empty
%</preamble>
\end{thebibliography}%


\begin{thebibliography}{99}
\bibitem{Falcon2010}E. Falcon, Discret. Contin. Dyn. Syst. B {\bf 13}, 819 (2010)
\bibitem{Newell2011}A. Newell and B. Rumpf, Annu. Rev. Fluid Mech. {\bf 43}, 59 (2011)
\bibitem{Residori2012}J. Laurie, U. Bortolozzo, S. Nazarenko, and S. Residori, Physics Report {\bf 514}, 121 (2012)

\bibitem{Zakharov1992}V. E. Zakharov, V. L'vov, and G. Falkovich, {\em Kolmogorov Spectra of Turbulence I} (Springer-Verlag, 1992); S. Nazarenko, {\em Wave Turbulence} (Springer, Berlin, 2010).

\bibitem{Levinsen00}E. Henry, P. Alstr{\o}m and M. T. Levinsen, Europhys. Lett. {\bf 52} 27 (2000); M. Yu. Brazhnikov et al., Europhys. Lett. {\bf 58} 510 (2002)
\bibitem{Snouck2009}D. Snouck, M.-T Westra, and W. van de Water, Phys. Fluids {\bf 21}, 025102 (2009)
\bibitem{Xia2010}H. Xia, M. Shats, and H. Punzmann, EPL {\bf 91}, 14002 (2010)

\bibitem{Ochi1998}M. K. Ochi, {\em Ocean waves} (Cambridge University Press, 1998)
\bibitem{Toba73}Y. Toba, J. Ocean Soc. Jpn. {\bf 29}, 109 (1973); K. K. Kahma, J. Phys. Oceanogr. {\bf 11}, 1503 (1981); G. Z. Forristall, J. Geophys. Res. {\bf 86} 8075 (1981); M. A. Donelan {\em et al.}, Philos. Trans. R. Soc. London A  {\bf 315}, 509 (1985); P. A. Hwang {\em et al.}, J. Phys. Oceanogr.  {\bf 30}, 2753 (2000)
\bibitem{Falcon2007}E. Falcon, C. Laroche, and S. Fauve, Phys. Rev. Lett. {\bf 98}, 094503 (2007)
\bibitem{Denissenko2007}P. Denissenko, S. Lukaschuk, and S. Nazarenko, Phys. Rev. Lett. {\bf 99}, 014501 (2007), S. Nazarenko, S. Lukaschuk, S. McLelland, and P. Denissenko, J. Fluid Mech. {\bf 642}, 395 (2010).
\bibitem{Cobelli2011}P. Cobelli, A. Przadka, P. Petitjeans, G. Lagubeau, V. Pagneux, and A. Maurel, Phys. Rev. Lett. {\bf 107}, 214503 (2011)

\bibitem{Royon07}A. Royon-Lebeaud, E. J. Hopfinger, and A. Cartellier, J. Fluid Mech. {\bf 577}, 467 (2007) and references therein.
\bibitem{Jalikop09}S. V. Jalikop and A. Juel, J. Fluid Mech. {\bf 640}, 131 (2009)

\bibitem{granular}K. Liffman, G. Metcalfe, and P. Cleary, Phys. Rev. Lett. {\bf 79}, 4574 (1997) S. G. K. Tennakoon, L. Kondic and R. P. Behringer, Europhys. Lett.  {\bf 45}, 470 (1999)

\bibitem{Falcon2008}E. Falcon, S. Auma\^{i}tre, C. Falc{\'o}n, C. Laroche and S. Fauve, Phys. Rev. Lett. {\bf 100}, 064503 (2008)
\bibitem{Lamb}H. Lamb, {\it Hydrodynamics} (Cambridge University Press, London, 6th Ed., 1975)
\bibitem{Deike12}L. Deike, M. Berhanu and E. Falcon, Phys. Rev. E {\bf 85}, 066311 (2012)

\bibitem{Zakharov1967}V. E. Zakharov, and N. N. Filonenko, Sov. Phys. Dokl. {\bf 11}, 881 (1967)
\bibitem{Zakharov1967-2}V. E. Zakharov, and N. N. Filonenko, J. App. Mech. Tech. Phys. {\bf 8}, 37 (1967)

\bibitem{Phillips1985} O. M. Phillips, J. Fluid. Mech. {\bf 4}, 426 (1958); A. C. Newell, and V. E. Zakharov, Phys. Lett. A {\bf 372}, 4230-4233 (2008)
\bibitem{Kuznetsov04}E. A. Kuznetsov, JETP Lett. {\bf 80}, 83 (2004)
\bibitem{Deike11}L. Deike, C. Laroche and E. Falcon, Europhys Lett. {\bf 96}, 34004 (2011).

\bibitem{Onorato2004}M. Onorato {\em et al.}, Phys. Rev. E {\bf 70}, 067302 (2004); M. Onorato {\em et al.}, Phys. Rev. Lett. {\bf 102}, 114502 (2009)
\bibitem{Falcon2011}E. Falcon and C. Laroche, EPL {\bf 95}, 34003 (2011)

\bibitem{Forristall2000}G. Z. J. Forristall, Phys. Oceanogr. {\bf 30}, 1931 (2000)
\bibitem{Socquet2005}H. Socquet-Juglard {\em et al.}, J. Fluid. Mech. {\bf 542}, 195 (2005)
\bibitem{Tayfun1980}M. A. Tayfun, J. Geophys. Res. {\bf 85}, 1548 (1980)

\end{thebibliography}
\end{document}